\documentclass{jfm}
\usepackage{graphicx}
\usepackage{amsmath}
\usepackage{tabularx}
\usepackage{color}
\usepackage{comment}
\usepackage[colorlinks,citecolor = blue, linkcolor=red,hyperindex,CJKbookmarks]{hyperref}

\usepackage{siunitx}
\newcolumntype{Y}{>{\centering\arraybackslash}X}
\newcolumntype{P}[1]{>{\centering\arraybackslash}p{#1}}

\shorttitle{Marangoni instability of different viscosities}
\shortauthor{Y. Li, J. G. Meijer and D. Lohse}

\title{Marangoni instabilities of drops of different viscosities in stratified liquids}

\author{Yanshen Li,\aff{1}
\corresp{\email{yanshen.li@utwente.nl}}
  Jochem G. Meijer,\aff{1}
  \and Detlef Lohse,\aff{1,2}
 \corresp{\email{d.lohse@utwente.nl}}
}

\affiliation{\aff{1}Physics of Fluids group, Max-Planck Center Twente for Complex Fluid Dynamics, Department of Science and Technology, Mesa+ Institute, and J. M. Burgers Centre for Fluid Dynamics, University of Twente, P.O. Box 217, 7500 AE Enschede, The Netherlands
\aff{2}Max Planck Institute for Dynamics and Self-Organization, 37077 G\"ottingen, Germany}

\begin{document}

\maketitle

\begin{abstract}
For an immiscible oil drop immersed in a stably stratified ethanol--water mixture, a downwards solutal Marangoni flow is generated on the surface of the drop, owing to the concentration gradient, and the resulting propulsion competes against the downwards gravitational acceleration of the heavy drop. In prior work of \citet{li2021Marangoni}, we found that for drops of \textit{low} viscosity, an oscillatory instability of the Marangoni flow is triggered once the Marangoni advection is too strong for diffusion to restore the stratified concentration field around the drop. Here we experimentally explore the parameter space of the concentration gradient and drop radius for \textit{large} oil viscosities and find a different and new mechanism for triggering the oscillatory instability in which diffusion is no longer the limiting factor. For such drops of higher viscosities, the instability is triggered when the gravitational effect is too strong so that the viscous stress cannot maintain a stable Marangoni flow. This leads to a critical drop radius above which the equilibrium is always unstable. Subsequently, a unifying scaling theory that includes both the mechanisms for low and for high viscosities of the oil drops is developed. The transition between the two mechanisms is found to be controlled by two length scales: The drop radius $R$ and the boundary layer thickness $\delta$ of the Marangoni flow around the drop. The instability is dominated by diffusion for $\delta<R$ and by viscosity for $R<\delta$. The experimental results for various drops of different viscosities can well be described with this unifying scaling theory. Our theoretical description thus provides a unifying view of physicochemical hydrodynamic problems in which the Marangoni stress is competing with a stable stratification. 

\end{abstract}

\section{Introduction}
Fluids in nature and technology are often multicomponent, with gradients in concentration when out of equilibrium \citep{lohse2020physicochemical}. The concentration gradients give rise to both density gradients and also solutal Marangoni stresses on an interface, thus leading to the competition between Marangoni convection and gravitational convection driven by density differences. Flows with such a competition are frequently encountered in technological applications. For example, in ink-jet printing, where the ink droplets are often multicomponent \citep{hoath2016fundamentals, lohse2022fundamental}; in freezing emulsions \citep{ghosh2008factors, degner2014factors, deville2017freezing, dedovets2018five}, where next to concentration gradients, thermal gradients also come into play; and in crystal growth \citep{chang1976analysis, schwabe1978experiments, schwabe1979some, chang1979thermocapillary, chun1979experiments, schwabe1982studies, preisser1983steady, kamotani1984oscillatory}, where the liquid column is subjected to a temperature gradient. The effect of gravity in these situations has often been ignored based on the perception that on small scales, gravitational effects are negligible owing to the small Bond number, which demonstrates the competition between gravity with capillary force \citep{nepomnyashchy2012interfacial}. However, \cite{edwards2018density}, \cite{li2019gravitational} and \cite{Diddens2021Competing} have recently discovered that gravitational effects can be important in evaporating binary droplets, in spite of the small scale reflected in small Bond numbers. It is therefore important to thoroughly understand the competing effect of gravity and Marangoni forces as some other flows where such a competition occurs may also need reconsideration.

The most basic example in which Marangoni forces compete with gravity is presumably an immiscible drop immersed in a stably stratified liquid, here made of an ethanol--water mixture.
Such a simple droplet system allows for a systematic investigation of the parameter space of the concentration gradient, the drop size and the drop viscosity. Without gravitational effects (i.e., without a density stratification), the concentration gradient only leads to simple motion of the immersed drop \citep{lagzi2010maze, maass2016swimming, jin2017chemotaxis, dedovets2018five}. However, for a stable stratification, owing to density differences and gravity, intriguing \& counter-intuitive phenomena can emerge, such as the oscillatory Marangoni flow around a drop/bubble \citep{zuev2006oscillation, viviani2008experimental, Schwarzenberger2015relaxation, Bratsun2018adaptive} or the continuous bouncing of a drop \citep{blanchette2012drops, li2019bouncing}. For low viscosity (\SI{5}{cSt}) drops, it has been found that the continuous bouncing is actually caused by an oscillatory instability \citep{li2021Marangoni} of the Marangoni flow, and the onset of the instability is triggered when the Marangoni advection is too strong for diffusion to restore the concentration field around the drop.

In this paper, we study the influence of the drop viscosity on the onset of the oscillatory instability in such a system by thoroughly exploring the three-dimensional parameter space spanned by the linearly stratified density gradient, the drop size and the drop viscosity. We start with very viscous silicone oil drops, for which we discover a new mechanism for the onset of the instability. We then develop a unifying scaling theory that includes both the newly discovered mechanism for high viscosity drops and the mechanism for low viscosity drops suggested by \cite{li2021Marangoni} and compare the results of this theory with the experimental results, which reveals good agreement. Finally, the instability thresholds for different viscosities are shown in one single phase diagram spanned by the Marangoni and Rayleigh numbers, thus clearly displaying the influence of the drop viscosity on the onset of the instability in the stratified flow.

\section{Experimental procedure \& methods}
\begin{figure}
\centering
\includegraphics[width=0.9\linewidth]{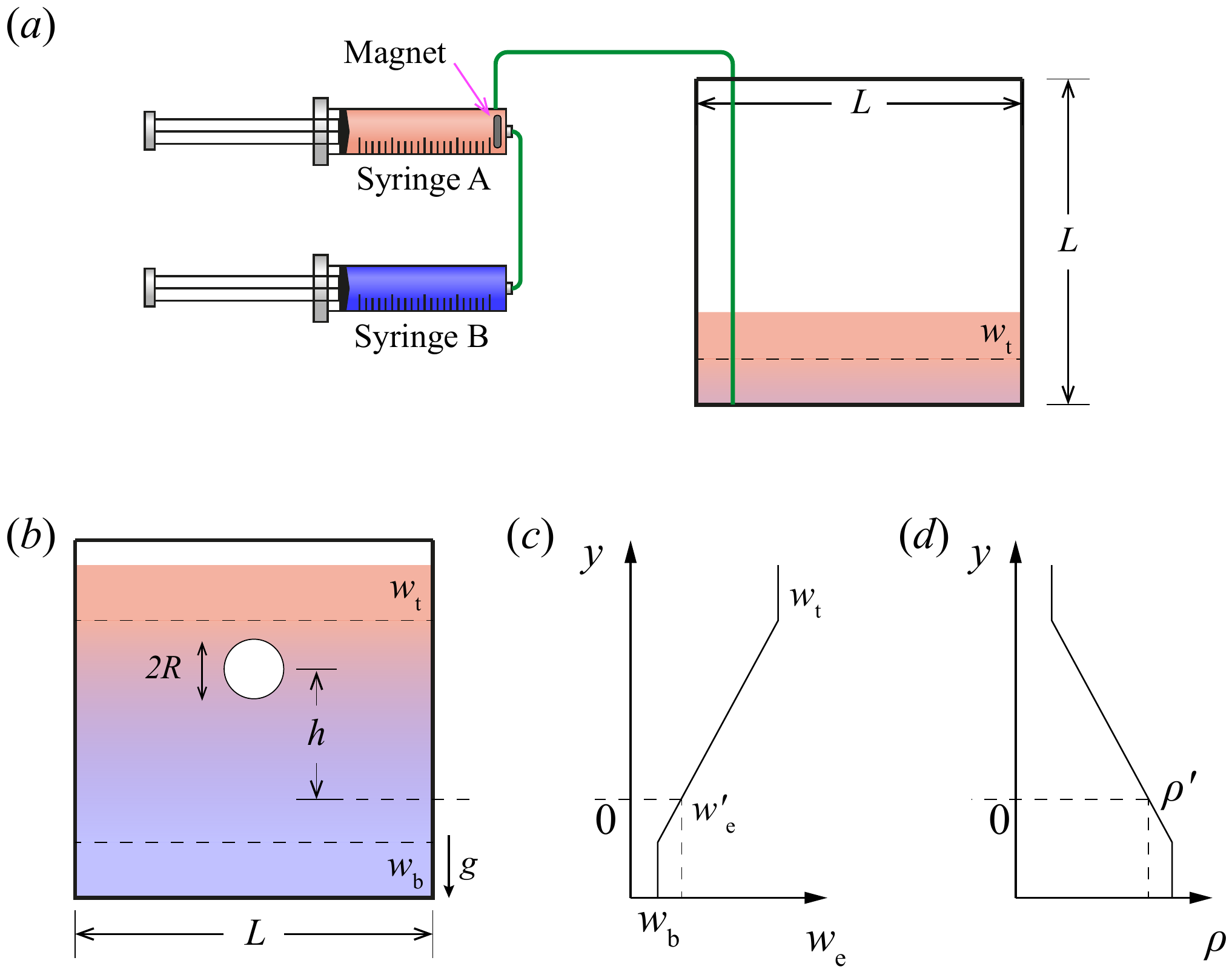}
\caption{Sketch of the experimental set-up. ($a$) The modified double-bucket method to generate the linearly stratified ethanol--water mixture in the glass container with inner length $L=\SI{30}{mm}$. Syringe A is filled with lighter liquid which has higher ethanol concentration, syringe B is filled with heavier liquid which has higher water concentration. The two syringes are filled to the same level and pumped at the same rate. During injection, the two liquids are mixed by a magnetic stirrer. The resulting mixture flows out of syringe A through a tube which leads to the bottom of the container. A lid which holds the tube in position is used to cover the container to prevent evaporation. A linear gradient is then formed. A liquid layer of uniform ethanol concentration $w_\mathrm{t}$ is first filled in the container before injection. ($b$) After the linear stratification has been formed, a layer of uniform ethanol concentration $w_\mathrm{b}$ is injected at the bottom. Red indicates the mixture is rich in ethanol and blue rich in water. Silicone oil drops of different radii $R$ and different viscosities $\nu^\prime$ are released in the stratified mixture. After releasing the drops, the container is again covered by a lid to prevent evaporation. Gravity is pointing downwards. ($c$) The ethanol weight fraction of the mixture $w_\mathrm{e}$ linearly increases from $w_\mathrm{b}$ at the bottom to $w_\mathrm{t}$ at the top. ($d$) The density of the mixture $\rho$ linearly decreases from the bottom to the top. The density of the silicone oil is $\rho^\prime$. The position $y=0$ marks the density matched position where $\rho(w_\mathrm{e}^\prime)=\rho^\prime$. The height of the drop is $h$, counting from $y=0$.}
\label{fig:1}
\end{figure}

Linearly stratified ethanol--water mixtures are prepared in a cubic glass container (Hellma, 704.001-OG, Germany) with inner length $L=\SI{30}{mm}$. Because the required liquid volume is very small ($<\SI{27}{mL}$), we use a slightly modified double-bucket method \citep{oster1965density} to generate the linear stratification, see figure \ref{fig:1}($a$). Syringe A is filled with lighter liquid which is rich in ethanol (with ethanol weight fraction $w_\mathrm{t}$), while syringe B is filled with heavier liquid which is rich in water. The two syringes are filled to the same level and mounted on one syringe pump (Harvard Apparatus, PHD 2000, USA) to make sure the two liquids are injected at the same rate. The injection rate is no larger than \SI{1}{mL/min}. During injection, the two liquids are constantly mixed in syringe A by a magnetic stirrer. The rod-like magnet in syringe A has a diameter of \SI{4}{mm}. The resulting mixture flows out of syringe A through a tube which leads to the bottom of the glass container. A linear gradient is formed during the process, see figure \ref{fig:1}($c$).  Strictly speaking, only equal-mass flux rate could lead to a perfectly linear gradient in weight fraction. The equal-volume flux rate in our case and the volume reduction during the mixing of ethanol and water both lead to an imperfect linear gradient. However, these imperfections are found to be very small and negligible, see figure \ref{fig:2}($e$). Especially, these imperfections are negligible as compared to the small sizes of the drops, thus we still call it linear in the following text. Notice that by this method, the linear gradient can be varied continuously. But in the experiments, we only choose several discrete values of the concentration gradient to span the parameter space.

Before injection, the container is first filled with a liquid layer of uniform ethanol concentration $w_\mathrm{t}$. After injection, a bottom layer of uniform ethanol concentration $w_\mathrm{b}$ is injected with a third syringe, so that the ethanol weight fraction of the mixture $w_\mathrm{e}$ increases continuously from $w_\mathrm{b}$ at the bottom to $w_\mathrm{t}$ at the top, see figure \ref{fig:1}($b$)\&($c$). The values of $w_\mathrm{b}$ and $w_\mathrm{t}$, and the depth of the linearly stratified layer are varied depending on the degree of stratification, which will be represented by $\mathrm{d}w_\mathrm{e}/\mathrm{d}y$ in the following text. To avoid bubble formation during mixing, both ethanol (Boom B.V., \SI{100}{\%}(v/v), technical grade, the Netherlands) and Milli-Q water are degassed in a desiccator at $\sim2000$ Pa for \SI{20}{min} before making the mixture. To reduce the disturbance to the linear gradient caused by the preferential evaporation of ethanol and subsequent Marangoni flows on the surface, the container is always covered by a lid except when releasing the drops. During injection, a lid which holds the tube in place is used. Otherwise, a glass lid is used to cover the container.

The ethanol weight fraction $w_\mathrm{e}(y)$ of the mixture at corresponding height $y$ is measured by laser deflection \citep{lin2013one, li2019bouncing} immediately after the stratified mixture is made. The two uniform layers $w_\mathrm{b}$ and $w_\mathrm{t}$ are used to increase the accuracy of the laser deflection method. The density of the mixture $\rho(y)$ is calculated from $w_\mathrm{e}(y)$ using an empirical equation \citep{Khattab2012density}, and $y=0$ is set to the position where the density of the mixture $\rho(0)$ equals that of the silicone oil $\rho^\prime$, see figure \ref{fig:1}($d$). The ethanol weight fraction at this density matched position $y=0$ is noted as $w_\mathrm{e}^\prime$. As reported previously \citep{li2021Marangoni}, the maximum extent of the flow field induced by the drop becomes larger for weaker density stratifications, so that the container may not be large enough for weaker density stratifications. However, since we are not interested in the effect of the finite size of the container, experiments for weak density stratifications, i.e., smaller concentration gradients $\mathrm{d}w_\mathrm{e}/\mathrm{d}y$, are repeated in a larger container (Hellma, 704.003-OG, Germany) with inner length $L=\SI{50}{mm}$. Results that are different for the two different container sizes are discarded, thus all the results presented in this paper are not influenced by the finite size of the container.

\begin{table}
\vspace{-0.25cm} 
	\begin{center}
	\begin{tabularx}{\textwidth}{p {3 cm} P {1.8 cm} P {1.8 cm} Y}
Liquid & Viscosity $\quad\quad\nu^\prime$ (\si{cSt}) & Density $\quad\quad\rho^\prime$ (\si{kg/m^3}) & Ethanol weight fraction $w_\mathrm{e}^\prime$ at the density matched position\\ [-3.5pt]
\hline \\[-12pt]
\SI{20}{cSt} silicone oil & 20 & 950 & \SI{30.2}{wt\%}\\
\SI{50}{cSt} silicone oil & 50 & 960 & \SI{24.5}{wt\%}  \\
\SI{100}{cSt} silicone oil & 100 & 966 & \SI{21.0}{wt\%} \\[-4pt]
\hline \\[-18pt]
\end{tabularx}
\caption{Properties of the silicone oils used in the experiments.}
\label{table1}
\end{center}
\vspace{-0.15cm} 
\end{table}

Silicone oils (Sigma-Aldrich, Germany) of different viscosities $\nu^\prime$ are used to generate drops of different radii $R$. They are injected from a \SI{1}{\micro L} syringe (Hamilton, KH7001) through a thin needle, whose outer-diameter is \SI{0.515}{mm}. The properties of the silicone oils are listed in table \ref{table1}. The drops are released from the top layer of the mixture, right after the laser deflection measurement. A collimated light-emitting diode (LED; Thorlabs, MWWHL4) is used to illuminate the drops in the mixture, and a camera (Nikon D850) connected to a long working distance lens system (Thorlabs, MVL12X12Z plus 0.25X lens attachment) is used to capture the side views of the drop. The radii and the trajectories of the drops are extracted from the sideview recordings. All images are recorded at 30 frames per second.

For each concentration gradient, a large drop is first released. After it has bounced 3 times, the drop is carefully taken out with another thin needle with outer-diameter \SI{0.515}{mm}, so as to minimize the disturbance to the stratified liquid. Then a second drop slightly smaller in size is released. In this way, we make sure that only one drop exists in the container at one time. This process is repeated until the smallest drop we could generate is released. One stratified mixture is used for no longer than \SI{40}{min}. Otherwise, a new stratified liquid with the same concentration gradient is generated again to continue the process. The data are analyzed after the first sweep in the drop radius $R$. If more data points are needed, the experiment is repeated at the same concentration gradient with the desired drop radii.

The flow field around the drop is revealed by adding tracer particles to the ethanol--water mixture. PSP (olyamid seeding particles, Dantec Dynamics, Denmark) particles of diameter \SI{20}{\micro\meter} are added in both ethanol and water at \SI{0.4}{mg/mL} before making the linearly stratified mixture. These particles follow the flow faithfully, see Supplemental Material for details. The interfacial tensions $\sigma$ between the silicone oils and the ethanol--water mixtures are measured by the pendant droplet method on a goniometer (OCA 15Pro, DataPhysics, Germany). Each measurement is repeated 6 times, and all the interfacial tension data are shown in figure \ref{fig:2}($c$). The solid lines are polynomial fits to the data points.

\section{Results for \SI{100}{cSt} silicone oil drops}

\begin{figure}
\centering
\includegraphics[width=1\linewidth]{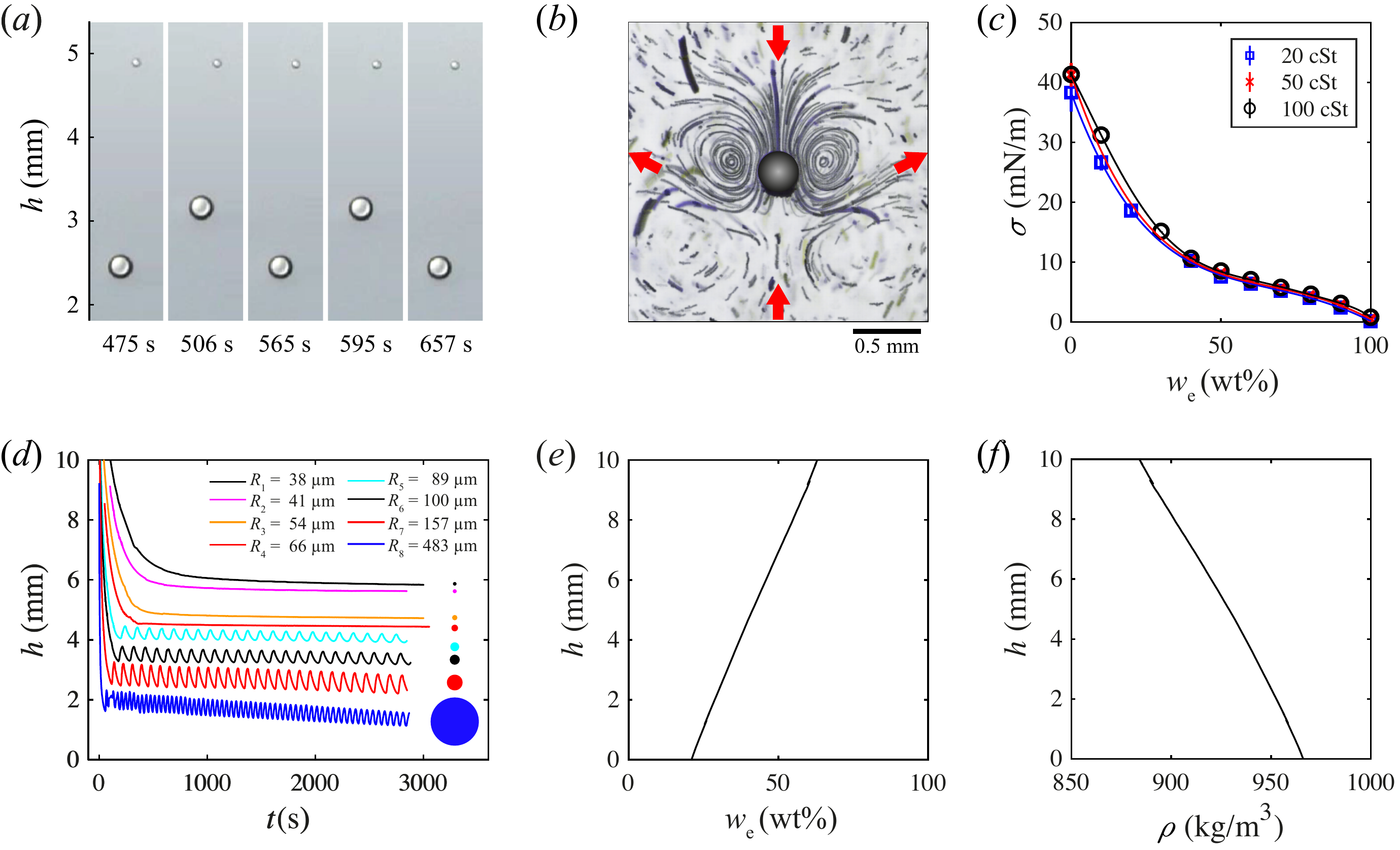}
\caption{($a$) Successive snapshots of two \SI{100}{cSt} silicone oil drops inside a linearly stratified ethanol--water mixture with $\mathrm{d}w_\mathrm{e}/\mathrm{d}y\approx\SI{40}{m^{-1}}$, at times after they are released. The smaller drop ($R=54\pm\SI{2}{\micro\meter}$) is levitating at $h\approx\SI{4.8}{mm}$, while the larger drop ($R=155\pm\SI{2}{\micro\meter}$) is bouncing at $h\approx\SI{3}{mm}$. ($b$) Stream lines around a levitating drop ($R=143\pm\SI{2}{\micro\meter}$) as revealed by the particle trajectories by superimposing 2300 images (covering \SI{76.7}{s}). The red arrows indicate the flow directions. The surrounding mixture has a smaller concentration gradient $\mathrm{d}w_\mathrm{e}/\mathrm{d}y\approx\SI{10}{m^{-1}}$. The scale bar is \SI{0.5}{mm}. ($c$) Interfacial tension $\sigma(w_\mathrm{e})$ between silicone oils of different viscosities and the ethanol--water mixture. Each point is an average of six measurements and the error bar is the standard deviation. The solid lines are polynomial fits to the data points. ($d$) Trajectories of \SI{100}{cSt} silicone oil drops of different radii in a linearly stratified ethanol--water mixture. The trajectories all start from $t = \SI{0}{s}$. Here, $h = 0$ is the density matched position, i.e., the position where $\rho=\rho^\prime$. The filled circles represent the relative sizes of the drops. Drops with radius $R$ smaller than \SI{66}{\micro\meter} are levitating, while drops with radius $R$ larger than \SI{89}{\micro\meter} are all bouncing. ($e$) The ethanol weight fraction $w_\mathrm{e}$ of the mixture at corresponding height $h$ is measured by laser deflection. ($f$) The density $\rho$ of the mixture at corresponding height is calculated from the ethanol weight fraction $w_\mathrm{e}$.}
\label{fig:2}
\end{figure}

For \SI{100}{cSt} silicone oil drops, two typical states are observed after the initial sinking phase, that is, levitating or bouncing. In figure \ref{fig:2}($a$), we show the successive snapshots of two \SI{100}{cSt} silicone oil drops in a mixture with $\mathrm{d} w_\mathrm{e}/\mathrm{d} y\approx\SI{40}{m^{-1}}$ (see also Supplementary Movie 1). While the smaller drop ($R=54\pm\SI{2}{\micro\meter}$) is levitating at $h\approx\SI{4.8}{mm}$, the larger drop ($R=155\pm\SI{2}{\micro\meter}$) is bouncing at $h\approx\SI{3}{mm}$. The value $h=0$ marks the position where the density of the oil ($\rho^\prime=\SI{966}{kg/m^3}$) equals that of the mixture (at $w_\mathrm{e}^\prime=\SI{21.0}{\%}$). The position of the drop $h$ as a function of time $t$ in the same stratified liquid, the ethanol weight fraction $w_\mathrm{e}$ and the density of the mixture $\rho$ at the corresponding height are respectively shown in figure \ref{fig:2}($d$), ($e$), and ($f$). The smallest drop ($R_1\approx\SI{38}{\micro\meter}$) is levitating at $h\approx\SI{5.9}{mm}$. As the drop size increases, it levitates at a lower position until, above a critical radius $R_\mathrm{cr}$, it starts to bounce instead of levitate. If its size is further increased, the drop bounces at a lower position (but still with $h>0$). 

The smaller drop is able to levitate above the density matched position ($h=0$) because of a stable Marangoni flow counteracting gravity, as revealed by the flow field around a levitating drop in a weaker gradient $\mathrm{d} w_\mathrm{e}/\mathrm{d} y\approx\SI{10}{m^{-1}}$, see the particle trajectories shown in figure \ref{fig:2}($b$) (and also Supplementary Movie 2). The interfacial tension of the drop $\sigma$ decreases with increasing ethanol concentration of the mixture $w_\mathrm{e}$, as shown in figure \ref{fig:2}($c$). The ethanol-concentration dependence of the interfacial tension of \SI{20}{cSt} and \SI{50}{cSt} silicone oils, which will be used later in the paper, are also shown in this figure. This interfacial tension gradient at the surface of the drop pulls liquid downwards, generating a viscous force acting against gravity, which levitates the drop.The lighter liquids that are pulled down then flow sideways because of buoyancy. When the drop becomes large enough, however, the equilibrium becomes oscillatory. The drop starts to bounce between two different levels (see Supplementary Movie 3). Notice that the onset of this oscillatory instability is essentially the instability of the flow field around a drop which is not moving. Once the drop starts to move (bounce in this case), the flow field becomes drastically different because now the translational motion of the drop must also be considered. Consequently, the analysis of the bouncing behaviour is completely different from that of the instability. In this study, we are only interested in the onset of the instability, and the transition from a levitating drop to a bouncing one serves as a good indication.

\begin{figure}
\centering
\includegraphics[width=0.45\linewidth]{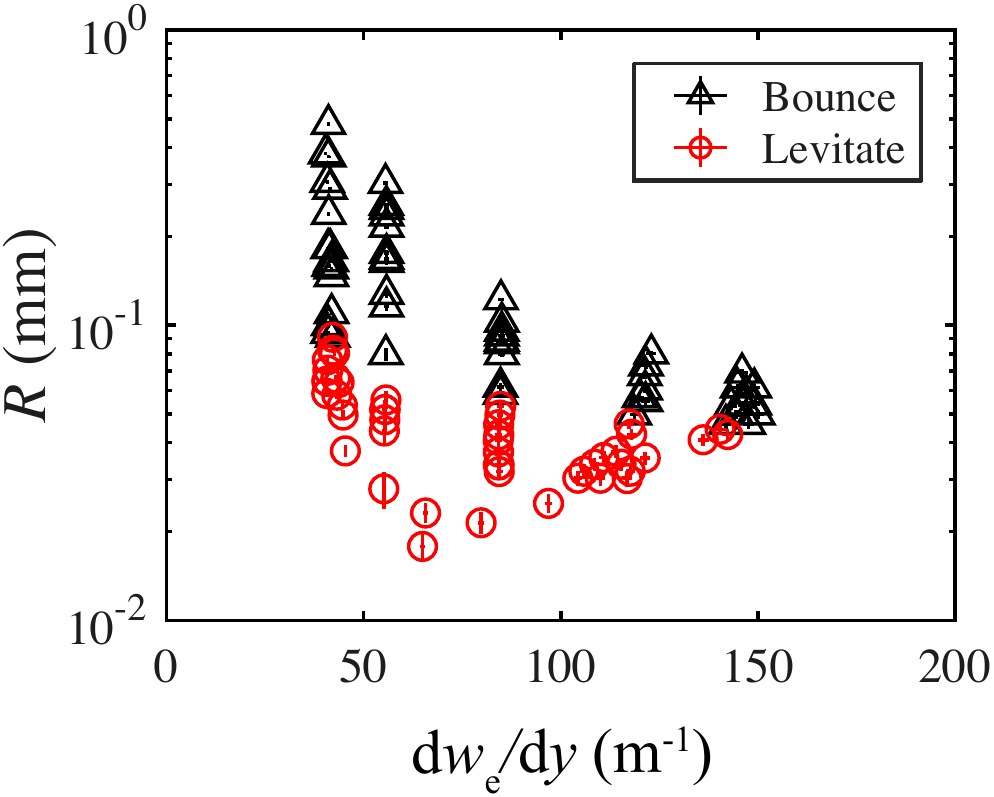}
\caption{Phase diagram of the \SI{100}{cSt} drops with a drop radius $R$ vs. concentration gradient $\mathrm{d}w_\mathrm{e}/\mathrm{d}y$ parameter space. Black triangles stand for bouncing drops, red circles for levitating ones. When experimentally determining the phase diagram, whether the drop bounces or not is decided within half an hour after its release, to avoid any untraceable change of the gradient arising from long-term mixing induced by the drop motion itself.}
\label{fig:3}
\end{figure}

Using this indication, we now explore the onset of this instability by independently varying $R$ and $\mathrm{d}w_\mathrm{e}/\mathrm{d}y$. Because we cannot vary the two parameters continuously in the experiments, we use the following criterion to determine whether a drop is bouncing: If the bouncing amplitude $h_\mathrm{A}$ of the drop is larger than its radius $R$, then the drop is considered to be bouncing (see Supplemental Material for more details). The results are shown in figure \ref{fig:3}. The parameter space for $\mathrm{d}w_\mathrm{e}/\mathrm{d}y\leq\SI{40}{m^{-1}}$ could not be explored because it is already contaminated by the finite size of the container. Different from the previously reported results of \SI{5}{cSt} silicone oil drops \citep{li2021Marangoni}, where there is a critical gradient $(\mathrm{d} w_\mathrm{e}/\mathrm{d} y)_\mathrm{cr}$ above which the Marangoni flow is always unstable, no such critical gradient is observed for \SI{100}{cSt} silicone oil drops up to $\mathrm{d}w_\mathrm{e}/\mathrm{d}y\approx\SI{150}{m^{-1}}$. On the contrary, for all the tested gradients, there is always a critical radius $R_\mathrm{cr}$ below which the flow is stable. This implies a different and new mechanism to trigger the Marangoni instability of \SI{100}{cSt} drops compared with that of the \SI{5}{cSt} drops. In the next section, we will develop a unifying scaling theory to account for both instability mechanisms, i.e., for low and high drop viscosity $\nu^\prime$.

\section{A unifying scaling theory for the onset of the instability}

\begin{figure}
\centering
\includegraphics[width=0.4\textwidth]{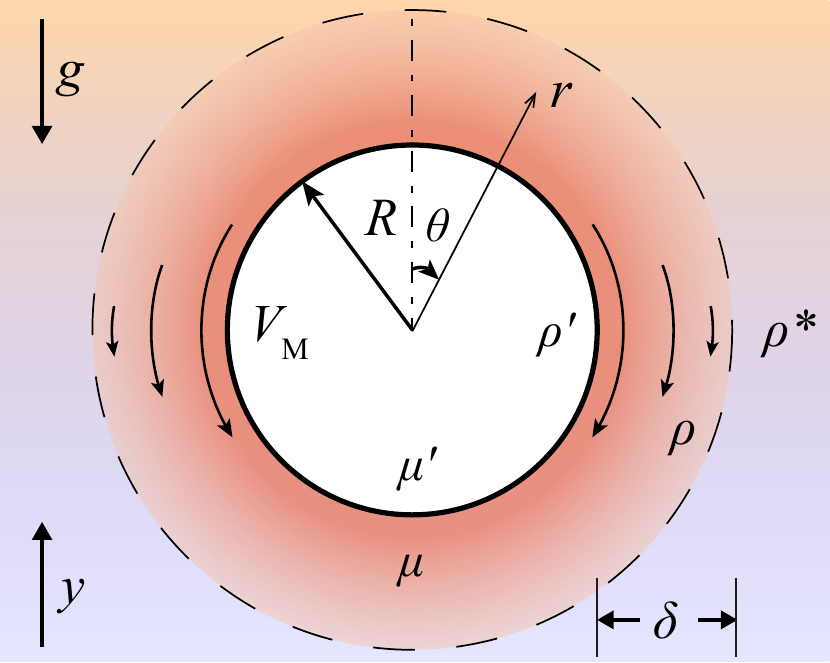}
\caption{A sketch of the levitating drop (of radius $R$, density $\rho^\prime$ and viscosity $\mu^\prime$) and the flow field and ethanol concentration around it. Deeper red means higher ethanol concentration. The shaded ring inside the dashed circle represents the kinematic boundary layer with thickness $\delta$ set by the Marangoni flow, as represented by the solid arrows. The ethanol concentration inside this layer is enhanced by Maragnoni advection bringing down the ethanol rich liquid. The Marangoni flow velocity at the equator of the drop is $V_\mathrm{M}$. The spherical coordinate $(r,\theta)$ has its origin at the center of the drop. Here $\rho$ and $\mu$ are the density and viscosity of the liquid inside this layer, and $\rho^*$ is the undisturbed density in the far field.}
\label{fig:sketch}
\end{figure}

For an immiscible drop immersed in the stably stratified ethanol--water mixture, a downwards Marangoni flow is generated on the surface of the drop, which also distorts the concentration field, see figure \ref{fig:sketch} for a sketch of the flow field and the concentration field around the drop. The system can be described in axisymmetrical spherical coordinates $(r, \theta)$ with its origin at the center of the drop. 

For the case of infinitely large solute diffusivity and zero density gradient, the analytical solution of the polar velocity $u_\theta$ outside the drop is given by \cite{young1959motion}:
\begin{equation}
u_\theta=-\frac{1}{4}\frac{1}{\mu+\mu^\prime}\frac{\mathrm{d} \sigma}{\mathrm{d} w_\mathrm{e}}\frac{\mathrm{d} w_\mathrm{e}}{\mathrm{d} y}R\left(\frac{R}{r}+\frac{R^3}{r^3}\right)\sin\theta,
\label{eq:youngUtheta}
\end{equation}
where $\mathrm{d}\sigma/\mathrm{d}w_\mathrm{e}$ is a material property (see figure \ref{fig:2}($c$)), $\mathrm{d}w_\mathrm{e}/\mathrm{d}y$ the undisturbed ethanol gradient of the mixture, $\mu^\prime$ the viscosity of the drop and $\mu$ the viscosity of the mixture. The Marangoni velocity at the equator of the drop is: 
\begin{equation}
u_\theta\vert_{r=R, \theta=90\si{\degree}}=-\frac{1}{2}\frac{\mathrm{d}\sigma}{\mathrm{d}w_\mathrm{e}}\frac{\mathrm{d}w_\mathrm{e}}{\mathrm{d}y}\frac{R}{\mu+\mu^\prime}.
\label{eq:youngVM}
\end{equation}
In our case, the ethanol diffusivity $D$ and the density gradient both have a finite value, so the influence of advection and buoyancy cannot be neglected, see figure \ref{fig:2}($b$). Especially, the Marangoni advection tends to homogenize the concentration field around the drop \citep{li2021Marangoni}, effectively decreasing the concentration gradient $\mathrm{d}w_\mathrm{e}/\mathrm{d}y$ around the drop, thus the Marangoni flow velocity is smaller than that predicted by Eq.(\ref{eq:youngVM}). Let $V_\mathrm{M}$ denote the Marangoni velocity at the equator of the drop in our case, i.e., $V_\mathrm{M}=u_\theta\vert_{r=R, \theta=90\si{\degree}}$, it is reasonable to drop the prefactor and write (also see Appendix \ref{appHeights}):
\begin{equation}
V_\mathrm{M}\sim-\frac{\mathrm{d}\sigma}{\mathrm{d}w_\mathrm{e}}\frac{\mathrm{d}w_\mathrm{e}}{\mathrm{d}y} \frac{R}{\mu+\mu^\prime}.
\label{eq:VM}
\end{equation}
The Marangoni flow imposes a viscous stress $\mu\nabla^2\mathbi{u}\vert_{r=R}$ on the liquid close to the drop, which, in the axisymmetrical spherical coordinate, is expanded as:
\begin{equation}
\mu\nabla^2\mathbi{u}\vert_{r=R}=\left.\mu\left[\frac{1}{r^2}\frac{\partial}{\partial r}\left(r^2\frac{\partial u_\theta}{\partial r}\right)+\frac{1}{r^2\sin\theta}\frac{\partial}{\partial\theta}\left(\sin\theta\frac{\partial u_\theta}{\partial\theta}\right)+\frac{2}{r^2}\frac{\partial u_r}{\partial\theta}-\frac{u_\theta}{r^2\sin^2\theta}\right]\right\vert_{r=R}.
\label{eq:UthetaExpand}
\end{equation}
For the liquid close to the equator of the drop, i.e., at $\theta=\SI{90}{\degree}$, $u_\theta\vert_{r=R, \theta=\SI{90}{\degree}}=V_\mathrm{M}$. Also notice that $u_r\vert_{r=R}=0$ and $\partial u_\theta/\partial \theta\vert_{\theta=\SI{90}{\degree}}=0$. The Marangoni viscous stress then becomes:
\begin{equation}
\mu\nabla^2\mathbi{u}\vert_{r=R, \theta=90\si{\degree}}\sim\mu\left(\frac{\partial^2V_\mathrm{M}}{\partial r^2}+\frac{2}{R}\frac{\partial V_\mathrm{M}}{\partial r}-\frac{V_\mathrm{M}}{R^2}\right).
\end{equation}
Let $\delta$ denote the boundary layer thickness of the Marangoni flow outside the drop, see figure \ref{fig:sketch}. Because $\partial V_\mathrm{M}/\partial r=\partial u_\theta/\partial r\vert_{r=R,\theta=\SI{90}{\degree}}\sim V_\mathrm{M}/\delta$ and $\partial^2 V_\mathrm{M}/\partial r^2=\partial u_\theta/\partial r\vert_{r=R,\theta=\SI{90}{\degree}}\sim -V_\mathrm{M}/\delta^2$, we obtain:
\begin{equation}
\mu\nabla^2\mathbi{u}\vert_{r=R, \theta=90\si{\degree}}\sim\mu\left(-\frac{V_\mathrm{M}}{\delta^2}+\frac{2V_\mathrm{M}}{\delta R}-\frac{V_\mathrm{M}}{R^2}\right).
\label{eq:ViscousPolar}
\end{equation}

Two length scales are present in Eq.(\ref{eq:ViscousPolar}), $\delta$ and $R$, and they are both in the denominator. The next question is, which one of them is the more relevant length scale, or in other words, which one of them is smaller. This can be inferred as follows: According to Eq.(\ref{eq:VM}), an increase in the drop viscosity $\mu^\prime$ would lead to a decrease in the Marangoni velocity $V_\mathrm{M}$. Consequently, the Reynolds number $Re=\rho V_\mathrm{M}R/\mu$ becomes smaller. Because within Prandtl--Blasius--Pohlhausen boundary layer theory \citep{schlichting2016boundary}, in general the boundary layer thickness $\delta$ is inversely proportional to the square root of the Reynolds number, then $\delta$ is larger. That is to say, an increase in $\mu^\prime$ would lead to an increase in $\delta$. Consequently, when $\mu^\prime$ is very small, it is possible that $\delta<R$ so that $\delta$ is the more relevant length scale. However, when $\mu^\prime$ is very large, it is possible that $\delta>R$ so that $R$ becomes the more relevant length scale. When $\mu^\prime$ is taking some intermediate value, it is possible that $\delta\sim R$ so that the situation is more complicated and we are entering a transitional case. The two limiting cases and the transitional case are considered subsequently in the following subsections.

\subsection{The limiting case when $\mu^\prime$ is very small}

When $\mu^\prime$ is very small, $\delta<R$ so that Eq.(\ref{eq:ViscousPolar}) reduces to
\begin{equation}
-\mu\nabla^2\mathbi{u}\vert_{r=R, \theta=90\si{\degree}}\sim\mu\frac{V_\mathrm{M}}{\delta^2}.
\label{eq:NSdelta}
\end{equation}
This is the downward viscous stress acting on the liquid close to the equator of the drop. This liquid is lighter than its surroundings as it is entrained from the top. The buoyancy force acting on it is $g\Delta\rho=g(\rho^*-\rho)$, where $\rho$ is its density, and $\rho^*$ is the density of the undisturbed liquid in the far field, see figure \ref{fig:sketch}. This liquid is brought down by the Marangoni flow along the surface of the drop, so $\Delta\rho\sim-R\,\,\mathrm{d}\rho/\mathrm{d}y$. Now let us look at the boundary layer which is thin, $\delta<R$, i.e., all the liquid in the boundary layer is close to the drop. The boundary layer experiences a downward viscous force $\mu V_\mathrm{M}/\delta^2$ and an upward buoyancy force $-gR\,\,\mathrm{d}\rho/\mathrm{d}y$. For a levitating drop, the two forces balance, thus
\begin{equation}
\mu\frac{V_\mathrm{M}}{\delta^2}\sim -gR\frac{\mathrm{d}\rho}{\mathrm{d}y}.
\label{eq:NS}
\end{equation}

\color{black}A very small $\mu^\prime$ also leads to a very large $V_\mathrm{M}$. This Marangoni advection acts to homogenize the concentration field inside the boundary layer while diffusion acts to restore it, see figure \ref{fig:sketch}. When advection is too strong, the concentration field close to the drop will be smoothened, thus weakening the Marangoni flow itself, then the flow becomes unstable. The advection time scale is $\tau_\mathrm{a}\sim R/V_\mathrm{M}$ and the diffusion time scale is $\tau_\mathrm{d}\sim \delta^2/D$. The flow will become unstable when advection is faster than diffusion, i.e.,
\begin{equation}
\tau_\mathrm{a}<\tau_\mathrm{d}.
\label{eq:9}
\end{equation}
Because diffusion is the limiting factor that causes this instability, we call it diffusion limited. Substituting the definitions of the two time scales into Eq.(\ref{eq:9}), we obtain
\begin{equation}
\frac{V_\mathrm{M}R}{D}>\frac{R^2}{\delta^2}.
\label{eq:10}
\end{equation}
The left-hand side of Eq.(\ref{eq:10}) has the form of a P\'eclet number, which is referred to as the Marangoni number 
\begin{equation}
Ma=\frac{V_\mathrm{M}R}{D}=-\frac{\mathrm{d}\sigma}{\mathrm{d}w_\mathrm{e}}\frac{\mathrm{d}w_\mathrm{e}}{\mathrm{d}y}R^2\,\frac{1}{(\mu+\mu^\prime)D}, 
\label{eq:Ma}
\end{equation}
where we have used Eq.(\ref{eq:VM}) with an equal sign. The instability criterion thus is
\begin{equation}
Ma>\frac{R^2}{\delta^2}.
\label{eq:DiffusionCriterion}
\end{equation}
Cancelling $\delta$ from Eqs. (\ref{eq:NS})\&(\ref{eq:DiffusionCriterion}), we obtain the diffusion-limited instability criterion for low viscosity $\mu^\prime$ cases:
\begin{equation}
Ma/Ra^{1/2}>s,
\label{eq:DiffusionThreshold}
\end{equation}
where 
\begin{equation}
Ra=-\frac{\mathrm{d}\rho}{\mathrm{d}y}\, \frac{gR^4}{\mu D}
\label{eq:Ra}
\end{equation}
is the Rayleigh number for characteristic length $R$ and $s$ is a constant to be determined. 

Eq.(\ref{eq:DiffusionThreshold}) can also be expressed in dimensional quantities by substituting the definitions of $Ma$ and $Ra$:
\begin{equation}
\frac{\mathrm{d}w_\mathrm{e}}{\mathrm{d}y}>\left(\frac{\mathrm{d}w_\mathrm{e}}{\mathrm{d}y}\right)_\mathrm{cr}=s^2\left(\mu+\mu^\prime\right)^2\frac{gD}{\mu}\frac{\mathrm{d}\rho}{\mathrm{d}\sigma}\frac{\mathrm{d}w_\mathrm{e}}{\mathrm{d}\sigma}.
\label{eq:Gradcr}
\end{equation}
Eq.(\ref{eq:Gradcr}) actually predicts a critical concentration gradient above which the flow is always unstable.

Eqs.(\ref{eq:DiffusionThreshold}) and (\ref{eq:Gradcr}) were confirmed by results of \SI{5}{cSt} silicone oil drops \citep{li2021Marangoni} and the instability threshold is measured to be $s=275\pm10$.

\subsection{The limiting case when $\mu^\prime$ is very large}

When $\mu^\prime$ is very large, $R<\delta$ so that Eq.(\ref{eq:ViscousPolar}) reduces to
\begin{equation}
-\mu\nabla^2\mathbi{u}\vert_{r=R, \theta=90\si{\degree}}\sim\mu\frac{V_\mathrm{M}}{R^2}.
\label{eq:ViscousR}
\end{equation}
Now the boundary layer is thick ($\delta>R$) and the force balance applied to the liquid close to the drop cannot be applied to the whole boundary layer. However, the liquid closest to the drop, or in other words, the most inner layer of the boundary layer, still experiences the very same two forces: The downward viscous force as shown by Eq.(\ref{eq:ViscousR}), and the upward buoyancy force $-gR\,\,\mathrm{d}\rho/\mathrm{d}y$. However, it also experiences a third force: An upward viscous force exerted by the outer layer. This is because the Marangoni velocity decreases radially outwards. Thus, for a stable Marangoni flow, we should have $\mu V_\mathrm{M}/R^2>-gR\,\,\mathrm{d}\rho/\mathrm{d}y$. Otherwise, the Marangoni flow will become unstable because the viscous force on the inner layer cannot overcome its buoyancy. The instability criterion thus is 
\begin{equation}
\mu\frac{V_\mathrm{M}}{R^2}< -gR\frac{\mathrm{d}\rho}{\mathrm{d}y}.
\label{neq:NSR}
\end{equation}
Because viscosity is the limiting factor that causes this instability, we call it viscosity limited. Substituting Eq.(\ref{eq:VM}) into Eq.(\ref{neq:NSR}) and reorganizing the left-hand side into the form of a Marangoni number, we obtain the viscosity-limited instability criterion
\begin{equation}
Ra/Ma>c,
\label{eq:ViscosityThreshold}
\end{equation}
where $c$ is a constant to be determined.

Eq.(\ref{eq:ViscosityThreshold}) can also be expressed in dimensional quantities by substituting the definitions of $Ma$ and $Ra$:
\begin{equation}
R>R_\mathrm{cr}=\displaystyle\sqrt{c\frac{\mu}{\mu+\mu^\prime}\frac{\mathrm{d}\sigma}{\mathrm{d}\rho}\frac{1}{g}},
\label{eq:Rcr}
\end{equation}
Eq.(\ref{eq:Rcr}) actually predicts a critical drop radius $R_\mathrm{cr}$ above which the flow is always unstable. The fact that the diffusivity $D$ does not enter into this equation further indicates that this instability criterion is not dominated by diffusion. Note that the concentration gradient $\mathrm{d}w_\mathrm{e}/\mathrm{d}y$ also does not enter into this equation, and the fluid properties $\mu$ and $\mathrm{d}\sigma/\mathrm{d}\rho$ depend on $w_\mathrm{e}$ -- the ethanol weight fraction at the levitation height -- thus $R_\mathrm{cr}$ is a function of $w_\mathrm{e}$.

Before moving on, we want to stress that Eq.(\ref{neq:NSR}) is a force imbalance that only applies to the liquid closest to the drop and it does not determine the levitation height of the drop. For the levitation height of the drop, please see Appendix \ref{appHeights}. It is also found that the instability threshold $c$ should increase with $\mu^\prime$, see Appendix \ref{appViscosityc}.

\subsection{The transitional case when $\mu^\prime$ is taking some intermediate value}

In between the two limiting cases, we may also have a transitional case where $\delta\sim R$ so that the two mechanisms are intertwined. However, $R$ is an independent variable. An increase in $R$ also leads to an increase in the Reynolds number, thus decreasing $\delta$. Consequently, upon increasing the drop radius $R$ from very small to very large, the onset of the instability would change from the viscosity-limited regime to the diffusion-limited regime. 

In the next section we will compare the experimental results of drops of different viscosities ($\nu^\prime=100$, 50 and \SI{20}{cSt}) with the different cases mentioned above and measure the corresponding instability threshold $c$. Because it is found that $c$ is different for oils of different viscosities, we use $c_{100}$, $c_{50}$ and $c_{20}$ in the following text to represent the respective instability thresholds for 100, 50 and \SI{20}{cSt} silicone oil drops. For general purposes, we will still use $c$, without specifying the drop viscosity. 

To calculate the Marangoni and Rayleigh numbers, ethanol weight fractions at the positions where the drops levitate $w_\mathrm{e}(h)$ are used to obtain the density $\rho$, viscosity $\mu$, diffusivity $D$ and the interfacial tension $\sigma$ (see Supplemental Material for the concentration dependence of $\rho$, $\mu$ and $D$). In the following, for bouncing drops, we use values corresponding to their highest positions. 

\section{Comparison with experimental results of different viscosities}

\begin{figure}
\centering
\includegraphics[width=0.45\linewidth]{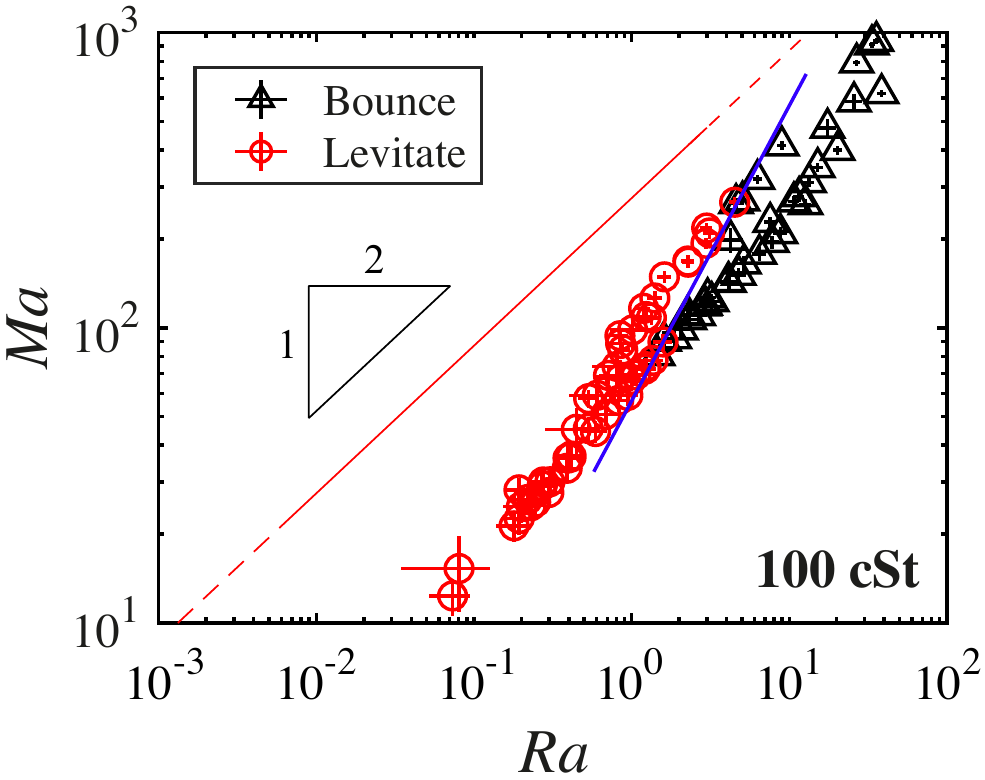}
\caption{Phase diagram of the \SI{100}{cSt} drops replotted in the $Ma$ vs. $Ra$ parameter space. Black triangles stand for bouncing drops, red circles for levitating ones. The red line is the diffusion-limited instability criterion $Ma/Ra^{1/2}=275$ for \SI{5}{cSt} silicone oil drops, below which the drops of low viscosity are levitating. The red solid line in the range $6\times10^{-3}\lesssim Ra\lesssim3$ has been confirmed by the experiments of \cite{li2021Marangoni} and the red dashed lines are not. The blue line separates the bouncing drops from the levitating ones for the high viscosity drops of this present paper.}
\label{fig:compare}
\end{figure}

\begin{figure}
\centering
\includegraphics[width=0.9\textwidth]{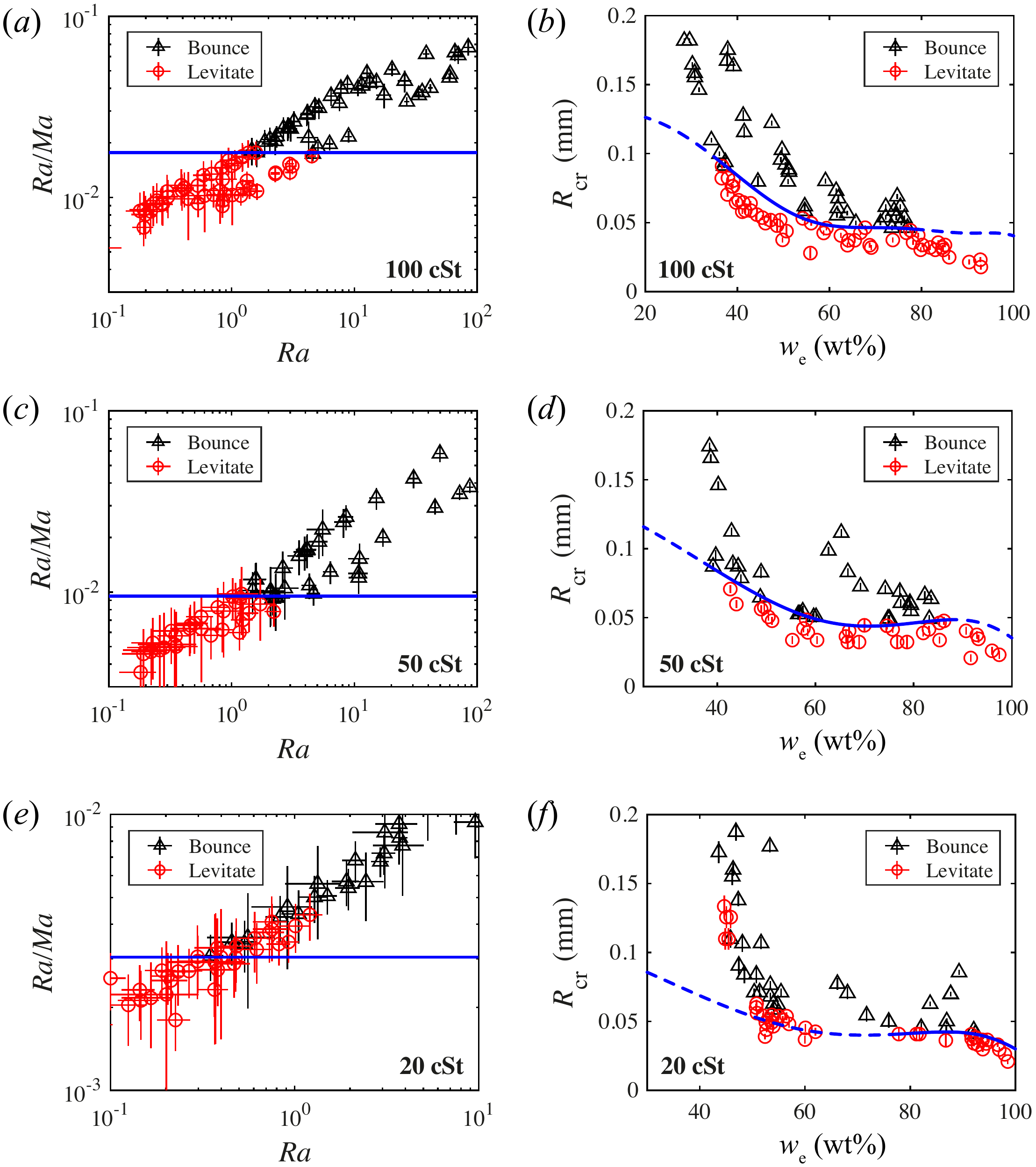}
\caption{Levitating/bouncing phase diagrams for silicone oils of different viscosities: 100, 50 and \SI{20}{cSt}. ($a$), ($c$), ($e$) $Ra/Ma$ vs. $Ra$ phase diagrams for 100, 50 and \SI{20}{cSt} silicone oils, respectively. As shown by the blue lines, the instability thresholds calculated by $(Ra/Ma)_\mathrm{cr}$ (see Eq.(\ref{eq:ViscosityThreshold})) are found to be $c_{100}\approx0.0177$, $c_{50}\approx0.0095$ and $c_{20}\approx0.003$, respectively. Black triangles stand for bouncing drops, and red circles for levitating ones. ($b$), ($d$), ($f$) Corresponding $R$ vs. $w_\mathrm{e}$ phase diagrams for 100, 50 and \SI{20}{cSt} silicone oils, respectively. Black triangles stand for bouncing drops and red circles for levitating ones. The blue lines are calculated $R_\mathrm{cr}$ with the corresponding instability thresholds from the left column of the figure. The blue solid lines in the range $\SI{36}{wt\%}<w_\mathrm{e}<\SI{77}{wt\%}$, $\SI{39}{wt\%}<w_\mathrm{e}<\SI{86}{wt\%}$ and $\SI{77}{wt\%}<w_\mathrm{e}<\SI{92}{wt\%}$ are those confirmed by experiments in each case.} 
\label{fig:6}
\end{figure}

The results of \SI{100}{cSt} silicone oil drops, as shown in figure \ref{fig:3}, are replotted in the $Ma$ vs. $Ra$ parameter space in figure \ref{fig:compare}. As a comparison, the instability threshold of the diffusion-limited case (for \SI{5}{cSt} silicone oil drops) $Ma/Ra^{1/2}\approx275$ is shown as the red line. In the diffusion-limited case, no drop should be bouncing below the red line. However, while all the data points of \SI{100}{cSt} drops are located below the red line, some of them still bounce. This indicates that diffusion is not the limiting factor to trigger the instability for the \SI{100}{cSt} drops. Moreover, the blue dashed line which separates the bouncing drops from the levitating ones has a slope close to one (see Eq.(\ref{eq:ViscosityThreshold})), which indicates that the instability is in the viscosity-limited regime. To further confirm this, we plot the results of \SI{100}{cSt} drops in the $Ra/Ma$ vs. $Ra$ parameter space in figure \ref{fig:6}($a$). As can be seen, there is indeed a critical value $(Ra/Ma)_\mathrm{cr}$ above which the Marangoni flow is unstable, and the instability threshold is measured to be $c_{100}\approx0.0177$ in the range $1\lesssim Ra\lesssim5$. 
The critical radius $R_\mathrm{cr}$ as a function of $w_\mathrm{e}$ with $c_{100}=0.0177$ is calculated from Eq.(\ref{eq:Rcr}) and shown as the blue curve in figure \ref{fig:6}($b$). The data shown in figure \ref{fig:6}($a$) are also replotted in figure \ref{fig:6}($b$) with ethanol weight fraction $w_\mathrm{e}$ measured at the levitating height as the $x$-axis. As can be seen, the blue curve nicely separates the levitating drops and the bouncing ones. The dashed blue line in the range $w_\mathrm{e}>\SI{77}{wt\%}$ ($w_\mathrm{e}<\SI{36}{wt\%}$) corresponds to $Ra<1$ ($Ra>5$) and is not confirmed by experiments. We cannot confirm the region for $Ra>5$ because this would require much larger levitating drops for which container wall effects could affect the data. We also cannot confirm the range for $Ra<1$, i.e., $w_\mathrm{e}>\SI{77}{wt\%}$, because for concentration gradient $\mathrm{d}w_\mathrm{e}/\mathrm{d}y$ larger than \SI{150}{m^{-1}}, the drops tend to levitate close to the top layer ($w_\mathrm{e}>\SI{77}{wt\%}$), and the concentration gradient in this region is influenced by diffusion because it is in contact with the uniform top layer (see figure \ref{fig:1}(b)): For $\mathrm{d}w_\mathrm{e}/\mathrm{d}y=\SI{150}{m^{-1}}$, the thickness of this region is $\Delta h\approx\SI{1.5}{mm}$. The diffusion time scale to change the concentration in this region is $\tau\sim \Delta h^2/D\approx\SI{37}{min}$. However, it normally takes $\sim\SI{20}{min}$ to make sure that a small drop is indeed levitating. During this time, the concentration in this region would have changed owing to diffusion, which makes the measurement not accurate. For lower concentration gradients, the drops levitate in the middle of the linear region that is farther from the top layer. The concentration field in this region is not influenced by diffusion. Though in figure \ref{fig:6}($a$) only half a decade in the range of $Ra$ is confirmed by the experimental results, the confirmed range in figure \ref{fig:6}($b$) covers more than half of the available range: $\SI{21}{wt\%}<w_\mathrm{e}<\SI{100}{wt\%}$ (because the drop cannot levitate below the density matched position $w_\mathrm{e}^\prime=\SI{21.0}{wt\%}$). 

The parameter spaces for \SI{50}{cSt} and \SI{20}{cSt} silicone oil drops are explored experimentally in the same way. Their interfacial tensions with the ethanol--water mixture are also measured in the same way, see figure \ref{fig:2}($c$). Their results are plotted in figure \ref{fig:6}($c$) \& ($d$) for \SI{50}{cSt} oil drops and figure \ref{fig:6}($e$) \& ($f$) for \SI{20}{cSt} oil drops. The instability threshold for \SI{50}{cSt} silicone oil is measured to be $c_{50}\approx0.0095$, and the blue line predicted from Eq.(\ref{eq:Rcr}) with $c_{50}=0.0095$ again nicely separates the levitating drops from the bouncing ones. The dashed blue line in the range $w_\mathrm{e}>\SI{86}{wt\%}$ ($w_\mathrm{e}<\SI{39}{wt\%}$) corresponds to $Ra<1$ ($Ra>2.5$) and is not confirmed by experiments owing to the same reason as above. Again, while only a small range of $Ra$ in figure \ref{fig:6}($c$) is confirmed by the experimental results, the confirmed range in figure \ref{fig:6}($d$) covers more than half of the available range: $\SI{25}{wt\%}<w_\mathrm{e}<\SI{100}{wt\%}$.

For \SI{20}{cSt} silicone oil drops, no constant instability threshold is found, see figure \ref{fig:6}($e$) \& ($f$), and we are probably entering the transitional case. Still, we cannot carry out experiments for $Ra<0.35$ and $Ra>16$ because of the same reasons as above. At $Ra\approx0.35$ ($\SI{77}{wt\%}<w_\mathrm{e}<\SI{92}{wt\%}$), the instability threshold is measured to be $c_{20}\approx0.003$, as shown by the black solid line in figure \ref{fig:6}($e$) and the blue solid line in figure \ref{fig:6}($f$). In the range $Ra>0.35$ ($w_\mathrm{e}<\SI{77}{wt\%}$), the instability threshold deviates from $c_{20}\approx0.003$.

\begin{figure}
\centering
\includegraphics[width=0.5\textwidth]{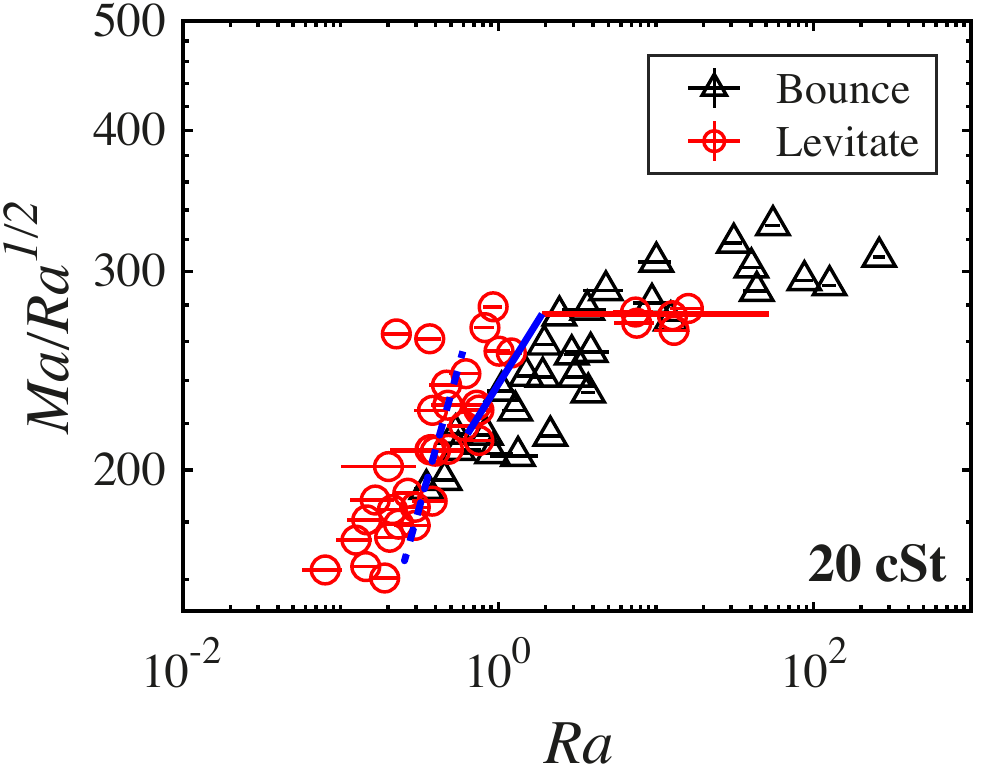}
\caption{Levitating/bouncing phase diagram for \SI{20}{cSt} silicone oil drops. The instability threshold starts from the viscosity-limited regime $Ra/Ma=0.003$ for $Ra<0.35$ (blue dashed line) and it slowly changes to $Ma/Ra^{1/2}=275$ for $Ra>3$ (red solid line). The instability threshold changes from the viscosity-limited regime to the diffusion-limited regime in between (blue solid line). The solid lines are confirmed by experiments and the dashed lines are inferred from the trend of the experimental results.} 
\label{fig:7}
\end{figure}

To get a better understanding of this case, results of \SI{20}{cSt} silicone oil drops are re-plotted in the $Ma/Ra^{1/2}$ vs. $Ra$ parameter space in figure \ref{fig:7}. The data points confirm that this is indeed the transitional case: The instability starts from the viscosity-limited regime $Ra/Ma=0.003$ when the drops are small ($Ra<0.35$), as shown by the blue dashed line. It slowly curves towards $Ma/Ra^{1/2}=275$ as the drop radius is increased ($0.35<Ra<3$) while still maintaining the viscosity-limited feature, as shown by the blue solid line. It finally turns into the diffusion-limited regime $Ma/Ra^{1/2}=275$ for much larger drops ($Ra>3$), as indicated by the red solid line.

We also performed experiments for \SI{10}{cSt} silicone oil drops, where the onset of the instability becomes diffusion limited and the instability threshold is measured to be $s=265\pm15$. This is indistinguishable from that of the \SI{5}{cSt} drops $s=275\pm10$ with the current experimental accuracy. We speculate that the instability threshold $s$ for the diffusion-limited case does not change with $\mu^\prime$, because the balance between advection and diffusion happens in the bulk and the boundary layer thickness $\delta$ in Eq.(\ref{eq:DiffusionThreshold}) has cancelled out. However, the instability threshold $c$ for the viscosity-limited case is indeed found to increase with $\mu^\prime$ (from $c_{20}\approx0.003$ to $c_{100}\approx0.0177$).

\section{Instability thresholds for different viscosities}

\begin{figure}
\centering
\includegraphics[width=0.5\textwidth]{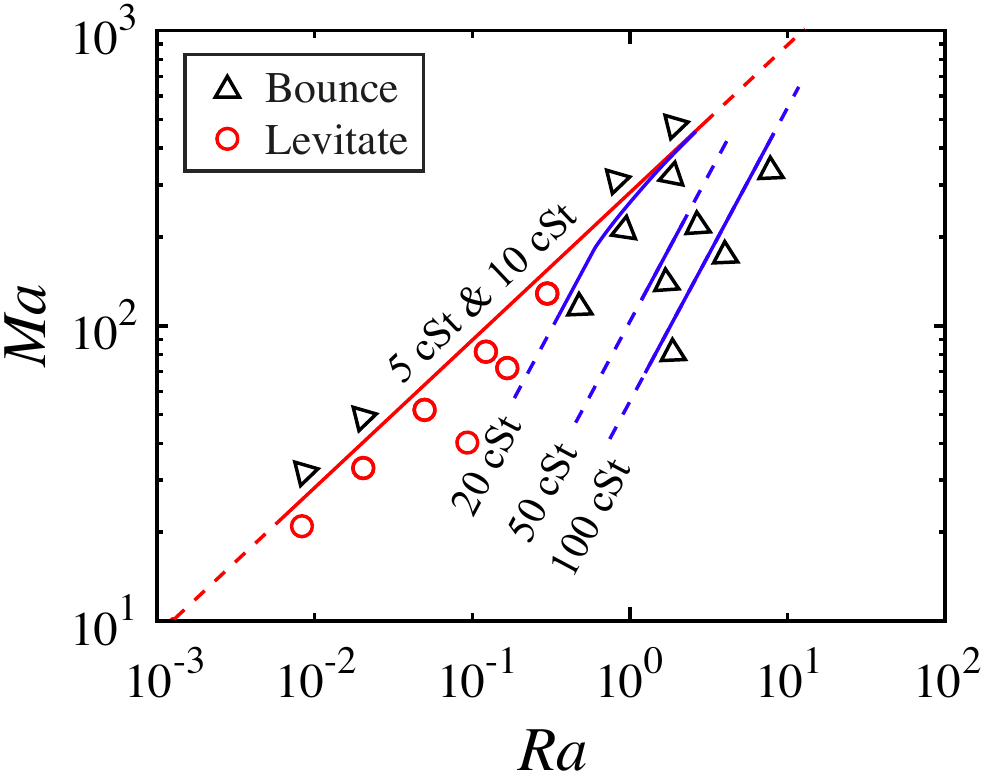}
\caption{Phase diagram of the instability thresholds for different drop viscosities. The red line is the diffusion-limited instability threshold $Ma/Ra^{1/2}=275$ for \SI{5}{cSt} silicone oil drops. Drops above the red line will be bouncing. The blue lines are the viscosity-limited instability thresholds, below which the drops will be bouncing. The instability threshold for \SI{20}{cSt} silicone oil changes from $Ma/Ra^{1/2}\approx275$ to $Ra/Ma\approx0.003$. The instability thresholds for \SI{50}{cSt} and \SI{100}{cSt} silicone oils are $Ra/Ma\approx0.0095$ and $Ra/Ma\approx0.0177$, respectively. Solid lines are confirmed by experiments, while the dashed lines are not.}
\label{fig:8}
\end{figure}

Now with all the results presented, we can have an overview of the instability thresholds for oils of different viscosities. They are plotted in figure \ref{fig:8}. When the oil viscosity is very low, the instability is diffusion limited, which yields $Ma/Ra^{1/2}=s$. This is the case for \SI{5}{cSt} and \SI{10}{cSt} silicone oil drops, and their instability thresholds are both $s\approx275$. When the oil viscosity is very high, the instability is viscosity limited, which yields $Ra/Ma=c$. This is the case for \SI{100}{cSt} and \SI{50}{cSt} silicone oil drops, and their instability thresholds are measured to be $c_{100}\approx0.0177$ and $c_{50}\approx0.0095$. When the oil viscosity is taking an intermediate value, the instability changes from the viscosity-limited regime to the diffusion-limited regime as the drop radius increases. This is the case for the \SI{20}{cSt} silicone oil drops, whose instability thresholds are measured to be $c_{20}\approx0.003$ and $s\approx275$.

\section{Conclusions and outlook}
In summary, the oscillatory instability of an immiscible oil drop immersed in a stably stratified ethanol--water mixture has been explored for oils of different viscosities. A unifying scaling theory has been developed, which predicts two different instability mechanisms depending on the two length scales in the system: The kinematic boundary layer thickness $\delta$ induced by the Marangoni flow, and the drop radius $R$. (i) When $\delta<R$, the instability is triggered when advection is too strong so that diffusion cannot restore the concentration field around the drop. This is the diffusion-limited regime which yields $Ma/Ra^{1/2}=s$. (ii) When $R<\delta$, the instability is triggered when the gravitational effect is too strong so that the viscous stress cannot maintain a stable Marangoni flow. This is the viscosity-limited regime which yields $Ra/Ma=c$. An increase in the drop viscosity leads to an increase in $\delta$, which eventually moves the instability criterion from the pure diffusion-limited regime to the pure viscosity-limited regime, through a transitional case which connects both regimes. Because $\delta$ cancels out in the diffusion-limited regime, the instability threshold $s$ does not change with an increase in the drop viscosity $\mu^\prime$, but the instability threshold $c$ in the viscosity-limited regime increases with increasing $\mu^\prime$. The scaling theory is well supported by the experimental results.

Owing to the limitations of the scaling theory, we do not know how $\delta$ quantitatively depends on the drop viscosity $\mu^\prime$ and drop radius $R$. Thus, we cannot predict beforehand in which regime will the instability be triggered. A quantitative understanding of $\delta$ may help to solve this problem. More detailed theoretical analysis may also be able to predict the instability thresholds beforehand. Nevertheless, our findings provide not only a deeper understanding to the simple droplet system under investigation, but also insight to other systems which allow for the direct competition between Marangoni and Rayleigh convection. For example, $Ra/Ma$ is called the dynamic Bond number in the context of Marangoni convections \citep{nepomnyashchy2012interfacial} $Bo_\mathrm{d}=Ra/Ma=\frac{\mu+\mu^\prime}{\mu}\frac{\mathrm{d}\rho}{\mathrm{d}\sigma}gR^2$. This number expresses the direct competition between Marangoni stress and gravity, as in the viscosity-limited regime. While the static Bond number $Bo_\mathrm{s}=\Delta\rho gR^2/\sigma$ determines whether a drop/bubble maintains a spherical shape, the dynamic Bond number $Bo_\mathrm{d}$ is more relevant for flows where the Marangoni flow is competing with a density gradient, such as in an evaporating binary sessile droplet \citep{edwards2018density, li2019gravitational, Diddens2021Competing}.

\begin{acknowledgements}
\section*{Acknowledgements}
We thank Vatsal Sanjay for valuable discussions. We acknowledge support from the Netherlands Center for Multiscale Catalytic Energy Conversion (MCEC), an NWO Gravitation programme funded by the Ministry of Education, Culture and Science of the government of the Netherlands, and D.L.'s ERC-Advanced Grant under project number 740479.
\end{acknowledgements}

\section*{Declaration of Interests}
The authors report no conflict of interest.

\appendix
\section{The levitation heights $h$ of the drops}\label{appHeights}

\begin{figure}
\centering
\includegraphics[width=0.6\linewidth]{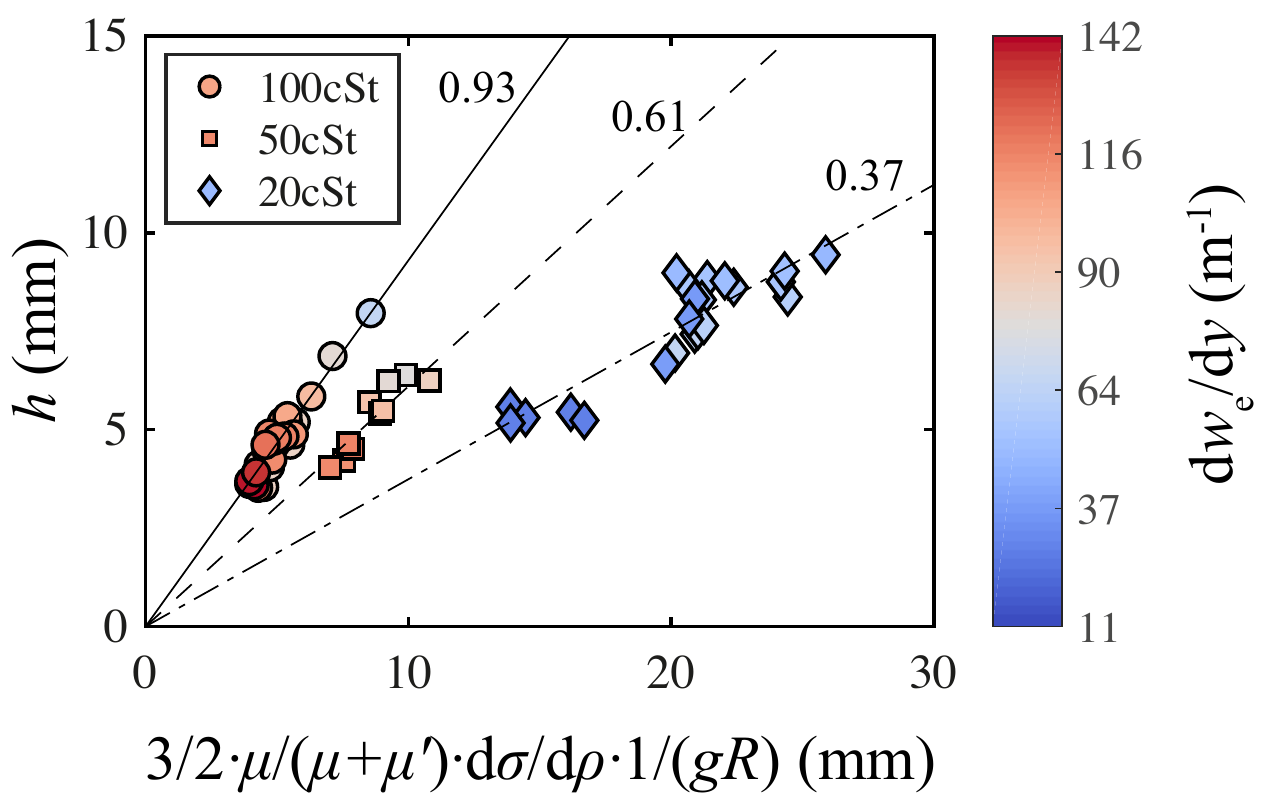}
\caption{Levitation heights $h$ of \SI{100}{cSt}, \SI{50}{cSt} and \SI{20}{cSt} silicone oil drops in linearly stratified ethanol--water mixtures with different concentration gradients $\mathrm{d}w_\mathrm{e}/\mathrm{d}y$. The prefactors $\alpha$ (defined in Eq.(\ref{si:eq:halpha})) are measured to be 0.93, 0.61 and 0.37 for 100, 50  and \SI{20}{cSt} silicone oil drops, respectively.}
\label{fig:heights}
\end{figure}

For a levitating drop, the viscous force imposed on it by the Marangoni flow is $F\sim\mu V_\mathrm{M}R$ and the gravity of the drop is $F_\mathrm{g}\sim gR^3(\rho^*-\rho^\prime)$. In a linear density gradient, $\rho^*-\rho^\prime=-h\cdot \mathrm{d}\rho/\mathrm{d}y$, where $h$ is the levitation height of the drop. Thus $F_\mathrm{g}\sim -gR^3h\cdot \mathrm{d}\rho/\mathrm{d}y$. From the force balance on the drop $F\sim F_\mathrm{g}$ we obtain:
\begin{equation}
\mu\frac{V_\mathrm{M}}{R^2}\sim-gh\frac{\mathrm{d}\rho}{\mathrm{d}y}.
\label{si:eq:hforce}
\end{equation}
Substituting the definition of $V_\mathrm{M}$ into Eq.(\ref{si:eq:hforce}), we can obtain the levitation height of the drop in a linear density gradient:
\begin{equation}
h\sim \frac{\mu}{\mu+\mu^\prime}\frac{\mathrm{d}\sigma}{\mathrm{d}\rho}\frac{1}{gR}.
\label{si:eq:hsim}
\end{equation}
For the case of infinitely large diffusivity and zero density gradient, the prefactor on the right hand side is found to be $3/2$ \citep{young1959motion}. But in our case, as mentioned earlier, the Marangoni advection acts to homogenize the concentration field around the drop, which decreases the upwards viscous force $F$. Thus we expect the prefactor to be smaller than $3/2$. We write this as
\begin{equation}
h=\alpha\cdot\frac{3}{2} \frac{\mu}{\mu+\mu^\prime}\frac{\mathrm{d}\sigma}{\mathrm{d}\rho}\frac{1}{gR}
\label{si:eq:halpha}
\end{equation}
with a prefactor $0<\alpha<1$, which reflects the strength of advection. A stronger advection will result in a smaller $\alpha$. Eq.(\ref{si:eq:halpha}) is confirmed by the levitation heights measured by drops of different sizes and viscosities, see figure \ref{fig:heights}. This also confirms that Eq.(\ref{eq:VM}) is the correct form to estimate $V_\mathrm{M}$. Finally, the measured $\alpha$ indeed becomes smaller for lower viscosities. 

\section{The influence of the drop viscosity $\mu^\prime$ on the viscosity-limited instability threshold $c$}\label{appViscosityc}

The fact that $\alpha$ is of the order of one indicates that the two sides of Eq.(\ref{si:eq:hforce}) are of the same order. To stress this, we rewrite Eq.(\ref{si:eq:hforce}) as
\begin{equation}
\mu\frac{V_\mathrm{M}}{R^2}\approx-gh\frac{\mathrm{d}\rho}{\mathrm{d}y}.
\label{si:eq:hforceapprox}
\end{equation}
At the onset of the viscosity-limited instability, see Eq.(\ref{neq:NSR}), we can write
\begin{equation}
\mu\frac{V_\mathrm{M}}{R_\mathrm{cr}^2}= -kgR_\mathrm{cr}\frac{\mathrm{d}\rho}{\mathrm{d}y}.
\label{si:eq:NSRk}
\end{equation}
where $k$ is a prefactor to be determined.
Combining Eqs.(\ref{si:eq:hforceapprox}) and (\ref{si:eq:NSRk}), one gets
\begin{equation}
h_\mathrm{cr}\approx kR_\mathrm{cr}.
\label{si:eq:hR}
\end{equation}
Because $h_\mathrm{cr}>\SI{3}{mm}$ (see figure \ref{fig:heights}) and $R_\mathrm{cr}<\SI{0.1}{mm}$ (see figure \ref{fig:3}), we have $k\gg1$. Comparing Eq.(\ref{si:eq:NSRk}) and Eqs.(\ref{neq:NSR}) and (\ref{eq:ViscosityThreshold}), one realizes that the instability threshold $c=1/k$, thus $c\ll1$. This is consistent with the experimental results. 

If the drop viscosity $\mu^\prime$ is increased while all other quantities remain the same, the levitation height $h$ will decrease according to Eq.(\ref{si:eq:halpha}). Then $k$ decreases according to Eq.(\ref{si:eq:hR}), thus $c$ increases. This explains why the instability threshold $c$ increases with increasing $\mu^\prime$.

\bibliographystyle{jfm}
\bibliography{References}

\end{document}